\documentclass[aps,prl,groupedaddress, showpacs, letterpaper, notitlepage,twocolumn]{revtex4-1}
\usepackage{graphicx}
\usepackage{dcolumn}
\usepackage{bm}
\usepackage{verbatim}
\usepackage{epstopdf}
\usepackage{amsmath}
\usepackage{amssymb}
\usepackage{amsthm}
\usepackage{natbib}
\usepackage{hyperref}
\usepackage{color}
\usepackage{comment}

\theoremstyle{definition}

\theoremstyle{remark}

\newcommand{\ket}[1]{\left| #1 \right\rangle}

\begin{document}

\title{Improved error thresholds for measurement-free error correction}
\date{\today }

\begin{abstract}
Motivated by limitations and capabilities of neutral atom qubits, we examine
whether measurement-free error correction can produce practical error
thresholds. We show that this can be achieved by extracting redundant
syndrome information, giving our procedure extra fault tolerance and
eliminating the need for ancilla verification. The procedure is particularly
favorable when multi-qubit gates are available for the correction step.
Simulations of the bit-flip, Bacon-Shor, and Steane codes indicate that
coherent error correction can produce threshold error rates that are on the
order of $10^{-3}$ to $10^{-4}$---comparable with or better than
measurement-based values, and much better than previous results for other
coherent error correction schemes.  This indicates that coherent error
correction is worthy of serious consideration for achieving protected
logical qubits.
\end{abstract}

\pacs{03.67.Pp,32.80.-t,32.80.Qk}
\author{Daniel Crow}
\author{Robert Joynt}
\author{M. Saffman}
\affiliation{
Department of Physics,
University of Wisconsin-Madison, 1150 University Avenue,  Madison, Wisconsin 53706}
\maketitle

An important near-term goal in quantum information processing is the
construction and operation of a high-quality logical qubit. This goal is
currently being pursued in several physical systems \cite{Schindler2011,
Waldherr2014, Kelly2015, Riste2015}. One promising candidate system is an
array of neutral atoms held in optical or magnetic traps \cite{Ladd2010,
Saffman2010}. The quantum information is stored in atomic hyperfine clock
states. This system has several attractive features: each natural qubit is
identical, clock states exhibit long coherence times measured in seconds,
and state preparation and state measurement can be performed on msec
timescales using well-developed techniques of optical pumping and detection
of resonance fluorescence \cite{Fuhrmanek2011,Gibbons2011}. Arrays of
individually addressable neutral atom qubits have been demonstrated in 1D 
\cite{Schrader2004,Knoernschild2010}, 2D \cite%
{Xia2015,Labuhn2014,Weitenberg2011,Schlosser2011}, and 3D \cite{YWang2015}.
Qubit numbers of order 100 have been demonstrated in 2D and 3D and in
principle these numbers could be extended to several thousands using
available technology. Lastly, the available gate set is universal, based on
microwave and laser light for single qubit rotations together with Rydberg
state mediated interactions for two-qubit, and multi-qubit, entangling gates 
\cite{Saffman2010}.

Achieving logical protection requires an error correction procedure
compatible with available operations. Standard error correction protocols
rely on performing frequent syndrome measurements \cite%
{nielsenchuang,Fowler2012}. This turns out not to be well suited for neutral
atom implementations for two reasons. First, the time needed for state
measurements is currently several orders of magnitude longer than for gate
operations. Second, it is difficult to measure a single atomic qubit in an
array without scattered light corrupting the state of nearby qubits,
although a number of possible solutions to this problem are under study \cite%
{Beterov2015}.

These challenges motivate the consideration of coherent, or
measurement-free, error correction (CEC) methods \cite%
{Nebendahl2009,Paz2010,tomita2012, Nebendahl2015}. Like standard
measurement-based error correction (MEC) \cite{Schindler2011,Waldherr2014,
Kelly2015, Riste2015}, techniques for measurement-free error correction are
based on stabilizer codes. However, there has been strong skepticism that
CEC can produce error thresholds close to those of MEC \cite%
{Aharonov2008,Shor1996,Steane1996}, though Paz-Silva \textit{et al.} did
achieve a CEC threshold only about one order of magnitude worse than MEC 
\cite{Paz2010}. We improve this result by nearly 2 orders of magnitude by
taking advantage of the resources available to neutral atoms, in combination
with a novel syndrome extraction technique.

CEC is particularly attractive for neutral atom and trapped ion approaches that rely on light scattering for entropy removal. 
As part of an error correction cycle, entropy in the data qubits is transferred
to fresh ancilla qubits, and is subsequently removed by
resetting the ancillas. Although an ancilla reset requires optical pumping and light scattering, the number of scattered photons is typically 1--2 orders
of magnitude less than would be needed for state measurement in MEC.

CEC can additionally benefit from an additional resource of neutral atom systems, since the computational capabilities 
include native Toffoli and C$_k$NOT gates. These
C$_k$NOTs can potentially achieve fidelities as high as 90\%
for  $k\sim 35$, while for smaller $k$ the fidelities of the native
gates are expected to beat fidelities of the decompositions
into 1- and 2-qubit gates \cite{Isenhower2011,Gulliksen2015}. Similarly, Rydberg interactions allow for parallel CNOT gates in which a single
control qubit targets multiple qubits simultaneously, improving the time required for syndrome extraction. Native Toffoli gates have also been demonstrated using trapped ion \cite{Monz2009} and superconducting qubits \cite{Wallraff2012}; thus the techniques presented
here could potentially be adapted to other platforms.

A quantum error correction code is determined by the number of physical
qubits and by the stabilizing group that fixes the logical subspace. This
stabilizer group, with elements $S_{i},$ is determined by its generators.
Given $n$ stabilizer generators, we can consider $2^{n}-1$ distinct
non-trivial products of the generators, forming additional stabilizers. If
stabilizer values could be extracted and processed without error, only the
stabilizer generators need to be measured, and additional stabilizers would
not provide additional useful information. The procedure we propose is to
copy onto ancillas the redundant information of a subset of these additional
stabilizers. This enables one not only to identify data errors, but also
errors that occur during syndrome extraction. The redundancy becomes useful
when combined with the Toffoli and C$_{k}$NOT gates, where the quantum gates
act as logical `AND' gates to ensure that stabilizer values agree,
conditionally targeting errors only if extracted stabilizer values match expected syndromes. 
Using this method, the ancilla qubits store only classical
information---i.e., they are immune to phase errors and are not directly
entangled with each other. We discuss this approach for three codes: 3-qubit
bit-flip, 9-qubit Bacon-Shor, and 7-qubit Steane.

\begin{figure}[h]
\includegraphics[width=.7\columnwidth]{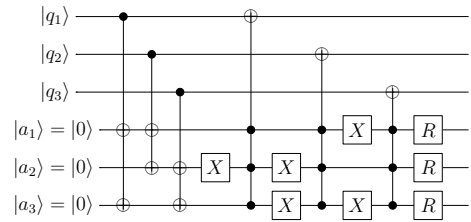}
\caption{The full measurement-free extraction and correction circuit for the
BF code. The first 3 gates are for syndrome extraction. The combination of $%
X $ gates and C$_{3}$NOT gates detect properly extracted syndromes and
correct errors accordingly. If a syndrome value is incorrectly extracted,
the data qubits are not affected. Reset operations are performed in the
final step, indicated with $R$ operations. This circuit also demonstrates the bit-flip correction procedure for the BS code, taking each $\ket{q_i}$ to be a row in the BS code. Then each CNOT gate is interpreted as 3 CNOT gates, one controlled by each qubit in the row. The C$_3$NOT gates target any single qubit in the row. A similar procedure is required for phase errors in the BS code.}
\label{fig:bitflip}
\end{figure}

The 3-qubit bit-flip (BF) code has logical states $\left\vert 0\right\rangle
_{L}=\left\vert 000\right\rangle $ and $\left\vert 1\right\rangle
_{L}=\left\vert 111\right\rangle $ with the usual stabilizers 
\begin{equation*}
S_{1}=Z_{1}Z_{2};\;\;\;S_{2}=Z_{2}Z_{3}.
\end{equation*}%
The values of $S_{1}$ and $S_{2}$ correctly identify single-qubit errors,
and each syndrome value corresponds to a distinct correction procedure.
Thus, any extraction errors leading to an incorrect value of either
stabilizer leads to an incorrect procedure, likely resulting in a logical
error. However, by considering the additional stabilizer 
\begin{equation*}
S_{3}=S_{1}S_{2}=Z_{1}Z_{3}
\end{equation*}%
it is possible to correctly identify if a single error occurs during ancilla
preparation or syndrome extraction (collectively: extraction errors). This
property follows from the fact that a correctly extracted syndrome always
produces an even number of ancillas in the logical $\left\vert
1\right\rangle $ state, as shown in Table \ref{table:syndromes}. Therefore,
a single extraction error occurs if an odd number of ancilla qubits occupy a
logical $\left\vert 1\right\rangle $ state. 
\begin{table}[!t]
\caption{Correctly extracted syndromes for single-qubit bit-flip errors on
the logical $\left\vert 000\right\rangle $ state. The table is easily
extended to errors on the $\left\vert 111\right\rangle $ state. }
\label{table:syndromes}\renewcommand\arraystretch{1.2} \setlength%
\tabcolsep{6pt}
\par
\begin{center}
\begin{tabular}{ccccc}
\hline\hline
& $\left| 000 \right\rangle$ & $\left| 100 \right\rangle$ & $\left| 010
\right\rangle$ & $\left| 001 \right\rangle$ \\ \hline
$Z_1Z_2$ & 0 & 1 & 1 & 0 \\ 
$Z_2Z_3$ & 0 & 0 & 1 & 1 \\ 
$Z_1Z_3$ & 0 & 1 & 0 & 1 \\ \hline\hline
\end{tabular}%
\end{center}
\end{table}
The error-correction circuit is shown in Fig. \ref{fig:bitflip}. The
circuit makes use of C$_{3}$NOT gates, to correct errors on the data qubits
only if the ancillary state corresponds to a valid syndrome. 

An advantage of using additional stabilizer information is that the
procedure does not require separate ancilla verification. That is,
single-qubit extraction errors can be detected simply from the combinatorics
of properly extracted syndromes. In our CEC circuits, the C$_k$NOT gates act
nontrivially on data qubits---i.e. correct errors---only if syndromes are
properly extracted. This implies that pre-existing data errors can survive a
faulty CEC cycle. However, with high probability the surviving data error is
simply corrected during the following cycle.

The 9-qubit Bacon-Shor (BS) code is obtained by layering the bit-flip code,
with one layer designed to protect against phase errors, resulting in a code
that can correct arbitrary single-qubit errors. The logical $X$ and $Z$
operators are just $X^{\otimes 9}$ and $Z^{\otimes 9}$, respectively. The
error correction procedure is quite similar to the bit-flip code, still
needing just 3 ancillas. Due to the underlying symmetry, this code requires
only 4 stabilizer generators. With the data qubits in a $3\times 3$ grid, the stabilizers are: 
\begin{align*}
Z_{U}=%
\begin{pmatrix}
Z & Z & Z \\ 
Z & Z & Z \\ 
I & I & I%
\end{pmatrix}%
,& \;\;\;Z_{D}=%
\begin{pmatrix}
I & I & I \\ 
Z & Z & Z \\ 
Z & Z & Z%
\end{pmatrix}%
, \\
X_{L}=%
\begin{pmatrix}
X & X & I \\ 
X & X & I \\ 
X & X & I%
\end{pmatrix}%
,& \;\;\;X_{R}=%
\begin{pmatrix}
I & X & X \\ 
I & X & X \\ 
I & X & X%
\end{pmatrix}%
.
\end{align*}

The procedure to perform error correction then proceeds in a manner similar
to the BF code. To correct bit-flip errors, we consider the additional
stabilizer $Z_{U}Z_{D}$. The circuit then proceeds as in Fig. \ref%
{fig:bitflip} but now taking each $\left\vert q_{i}\right\rangle $ to
correspond to a single row of 3 data qubits. Each CNOT gate in the circuit
can then be interpreted as 3 physical CNOT gates---one for each data qubit.
The C$_{k}$NOT gates can target any single physical qubit in the row. The
procedure for correcting phase errors is analogous, although extraction and
correction is done by column. Additional information is provided in the supplementary material. 

The 7-qubit Steane code has 6 stabilizer generators and requires 7 ancillas
to correct arbitrary single-qubit errors. Three $Z$-type stabilizers 
\begin{equation*}
S^Z_1 = Z_1Z_2Z_3Z_7; \;\;\; S^Z_2 = Z_1Z_2Z_4Z_6; \;\;\; S^Z_3 =
Z_1Z_3Z_4Z_5
\end{equation*}
detect bit-flip errors, while $X$-type stabilizers detect phase-flip errors
and are obtained from the $Z$-type operators by replacing each $Z_i$ with $%
X_i$. The logical operators are $Z_L = Z^{\otimes 7}$ and $X_L = X^{\otimes
7}$

\begin{figure}[h]
\begin{center}
\includegraphics[width=\columnwidth]{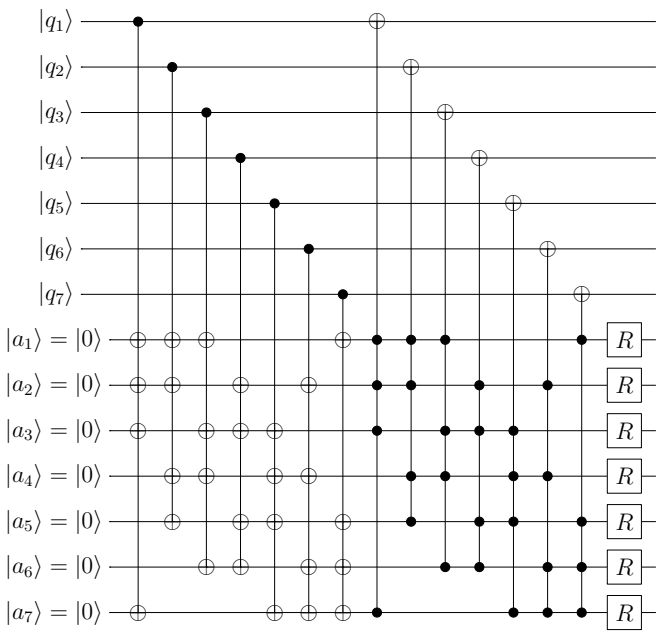}
\caption{Error correction circuit for the Steane code for bit-flip errors. The circuit for phase errors is similar.}
\label{fig:steane-circuit}
\end{center}
\end{figure}

We will restrict our discussion to bit-flip errors; phase errors follow
analogously. With 3 $Z$-type generators, we can form 7 distinct stabilizers.
Then error correction proceeds as follows: (1) extract the 7 stabilizer
values onto 7 ancilla qubits, and (2) use a sequence of 7 C$_4$NOT gates to
correct errors, matching each target data qubit $q_i$ to the unique set of
control ancilla qubits whose corresponding stabilizers act on $q_i$. The
details of this procedure are discussed in the supplementary material, and the circuit is shown in Fig. \ref{fig:steane-circuit}.

We performed a numerical simulation of measurement-free error correction
using the circuits shown in the previous section. We adopted an error model
controlled by two error-rate parameters: the gate rate $p_{\mathrm{gate}}$,
and the memory (or idle-gate) rate $p_{\mathrm{mem}}$. All single-qubit gate
errors are all assumed to be depolarizing, i.e. if an error occurs on qubit $%
i$, then a single-qubit Pauli is selected at random and applied to qubit $i$%
. Two-qubit gate errors occur with the same probability $p_{\mathrm{gate}}$
as single-qubit gates, but the error is chosen at random from the set of
2-qubit Paulis. For multi-qubit gates, each control--target pair of qubits
is treated as a two-qubit gate site, subject to error model as other two qubit errors. The effect is that C$_k$NOT gates have an error rate of roughly $k\cdot p_{\mathrm{gate}}$, roughly matching physical error models \cite%
{Gulliksen2015}.  

The simulated circuits required the ability to perform single-qubit Pauli, CNOT, and C$_k$NOT gates. The state evolution
was performed using stabilizer simulation, in a manner similar to the
techniques outlined by Aaronson  and Gottesman in \cite{Aaronson2004}. However, the C$_k$%
NOT gate is not in the Clifford group, and is not typically simulable in an
efficient manner. However, in every circuit studied here, the C$_k$NOT gates
are always controlled by the ancilla qubits, which only store classical
information and are modeled as classical bits.

To efficiently collect data on the circuit, we used simulation and
computational techniques similar to those in Refs. \cite{Bravyi2013} and 
\cite{Wang2010}, with additional detail in the supplementary material. Using these techniques, we can easily and accurately
estimate logical error rates. In principle, these methods could be scaled to
more qubits and additional input parameters in a straightforward manner.

\begin{figure}[h]
\includegraphics[width=\columnwidth]{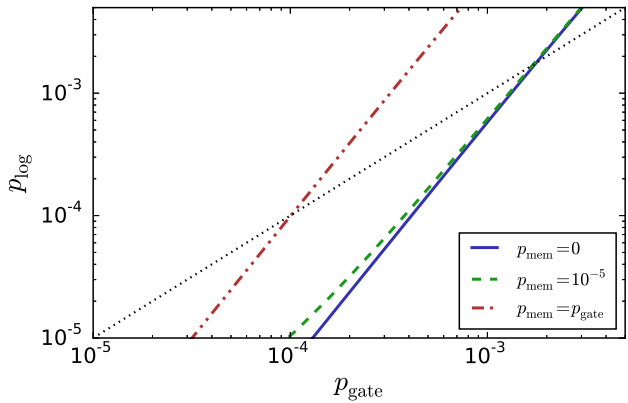}
\caption{(Color online) Logical error rate vs. gate error rate for the Bacon-Shor code,
with three different choices of memory error rate. The dotted line shows $p_{%
\mathrm{log}}=p_{\mathrm{gate}}$. The difference between the curves with
memory rates of 0 and $10^{-5}$ is minimal.}
\label{fig:shor-rates}
\end{figure}

\begin{figure}[tbp]
\includegraphics[width=\columnwidth]{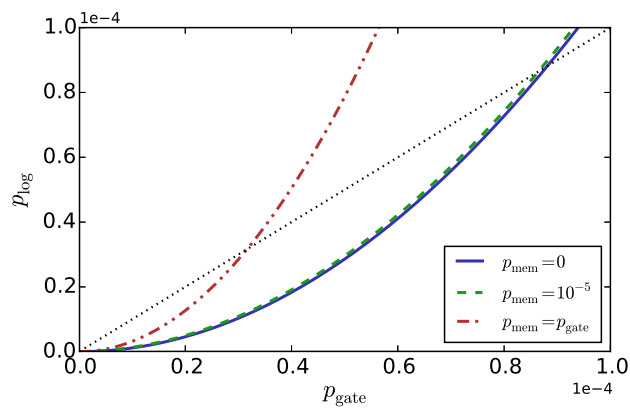}
\caption{(Color online) Logical error rate vs. gate error rate for the Steane code, with
three different choices of memory error rate. The dotted line shows $p_{%
\mathrm{log}}=p_{\mathrm{gate}}$. Note the near overlap between the curves
with memory rates of 0 and $10^{-5}$. }
\label{fig:cc-rates}
\end{figure}

The threshold was evaluated by determining $p_{\mathrm{gate}}$ such that the
logical error rate $p_{\mathrm{log}}$ satisfied $p_{\mathrm{log}}(p_{\mathrm{%
gate}})=p_{\mathrm{gate}}$. To reduce $p_{\mathrm{log}}$ to a function of a
single parameter, we set $p_{\mathrm{mem}}$ to a fixed value, or set $p_{%
\mathrm{mem}}=p_{\mathrm{gate}}$. For neutral atom qubits, memory error rates are
one to two orders of magnitude below gate rates. In this region of parameter
space, varying $p_{\mathrm{mem}}$ had little effect on the threshold gate
rate, demonstrated in Figs. \ref{fig:shor-rates} and \ref{fig:cc-rates}. The
threshold results are summarized in Table \ref{table:values}. 
\begin{table}[!tp]
\caption{Comparison of CEC and MEC gate thresholds. The BS and Steane MEC values are the
best values obtained for each code in Ref. \protect\cite{Cross2009}, while the BF value is obtained from \protect\cite{Silbey2005} and is scaled by 1.5 since our error rate includes phase errors.}
\label{table:values}\renewcommand\arraystretch{1.2} \setlength\tabcolsep{6pt}
\par
\begin{center}
\begin{tabular}{lccc}
\hline\hline
& BF & BS & Steane \\ \hline
CEC $p_{\mathrm{mem}}=0$ & $0.010$ & $1.8 \times 10^{-3}$ & $8.9 \times 10^{-5}$  \\ 
CEC $p_{\mathrm{mem}}=p_{\mathrm{gate}}$  & $5.5 \times 10^{-4}$ & $1.01 \times 10^{-4}$ & $3.2
\times 10^{-5}$ \\ 
MEC $p_{\mathrm{mem}}=0$ && $2.6 \times 10^{-4}$ & $5 \times 10^{-4}$  \\ 
MEC $p_{\mathrm{mem}}=p_{\mathrm{gate}}$ & $\sim 0.03$ & $2.1 \times 10^{-4}$ & $2.6
\times 10^{-4}$   \\ \hline\hline
\end{tabular}%
\end{center}
\end{table}


The difference between the thresholds for the Bacon-Shor and Steane codes
highlights the behavior of C$_k$NOT gates with unprotected ancilla qubits.
In the Steane code, the successful correction of a data error depends on the
successful extraction of 4 syndrome values, while the Bacon-Shor depends on
only 3 syndrome values. Furthermore, the syndrome extraction process for the
Steane code requires 56 CNOT gates, compared with 36 for the Bacon-Shor
code. Thus, the C$_k$NOT gates performing error correction are significantly
more likely to fail in the case of the Steane code. However, in all
procedures studied here, failures in extraction do not propagate \emph{new}
errors onto data qubits.

Earlier work on measurement free error correction found a threshold of $%
p_{T} \approx 3.8\times10^{-5}$ for the 9-qubit Bacon-Shor code \cite{Paz2010}. Thus, our
work indicates a substantial improvement over this value. Additionally, the
earlier value needs 18 additional ancilla qubits, while our protocol needs
just 3. The differences can be attributed to the combination of extracting
additional stabilizer values, coupled with the efficiency of C$_k$NOT gates
for performing classical logic.

Directly comparing our result to measurement-based results is not
straightforward -- measurement-based values depend on the chosen ancilla
verification scheme and do not use extra stabilizer information. In
addition, there is some arbitrariness in our choice of an error model for C$%
_{k}$NOT gates. With these caveats, in Table \ref{table:values}, we
compare our results to the best measurement-based threshold values from Ref. 
\cite{Cross2009}, which are also first-level depolarizing thresholds. The
dramatic difference in thresholds for the case of the Bacon-Shor seems to
exist only in the regime where memory error rates are small. In this regime,
errors are dominated by gate errors, but the circuit lengths for CEC using
neutral atom resources are typically quite small -- and certainly smaller
than those required for ancilla verification. Without the efficiency of
multiqubit resources, we would expect thresholds to drop.

Somewhat surprisingly, the thresholds calculated for CEC are comparable to,
and, in the case of Bacon-Shor, better than thresholds calculated for MEC. 
The threshold error rates are encouraging -- the bit-flip and Bacon-Shor
codes both yield values that are realistic for neutral atom systems \cite%
{Theis2016}. The overhead required for CEC is not greater, and possibly even
less than that in MEC, though C$_{k}$NOT gates are required. Furthermore,
the technique of using redundant syndrome extraction can potentially be
useful in other architectures. Certainly, measurement problems are not
restricted to neutral atoms. Furthermore, redundant syndrome extraction can
potentially be used in MEC to avoid ancilla verification, which we plan to
explore in future research.

The simulations were performed using the University of Wisconsin Center for
High Throughput Computing. We gratefully acknowledge the support of the
staff. MS acknowledges support from the IARPA MQCO program through
ARO contract W911NF-10-1-0347 and the ARL-CDQI through cooperative agreement W911NF-15-2-0061. We also thank G. A. Paz-Silva, S. N. 
Coppersmith, M. Friesen, J. Ghosh, and E. Ercan for helpful discussions and
communications.

\bibliographystyle{apsrev4-1}
\bibliography{bibliography.bib}

\widetext
 
\section*{Supplementary material }
\label{app:steane}

\subsection*{Additional details for the Steane code}

Here we provide further detail on the Steane code circuit, and the additional stabilizer information in particular. Since much of this discussion is relevant to measurement-based error correction, we will begin with that case before proceeding to the measurement-free case. The Steane code requires 7 data qubits, and has 6 stabilizer generators, encoding a single logical qubit. The $X$ and $Z$ logical operators are $X_L = X^{\otimes 7}$ and $Z_L = Z^{\otimes 7}$. The bit-flip and phase-flip errors can be treated independently in the Steane code, so we will restrict ourselves to the discussion of bit-flip errors. 

Here, to emphasize the symmetry of the Steane code, we adopt a different labelling convention for the stabilizers. For bit-flip errors, we will take the 3 $Z$-type stabilizers to be
\begin{align*}
S_1 &= Z_1Z_2Z_3Z_7, \\
S_2 &= Z_1Z_2 Z_4 Z_6, \\
S_3 &= Z_1 Z_3 Z_4 Z_5. \\
\end{align*}

\begin{figure}[h]
\begin{center}
\includegraphics[width=.3\textwidth]{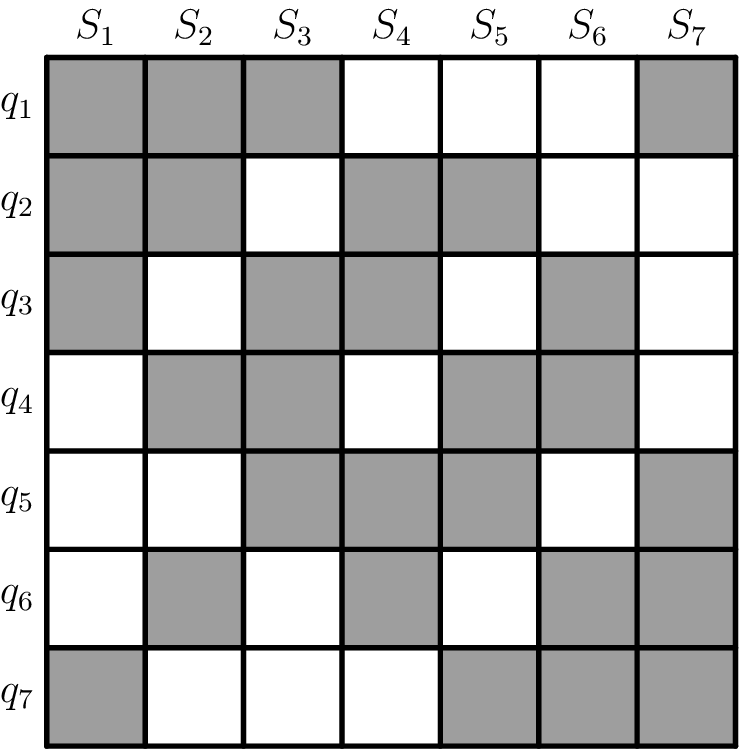}
\caption{The qubit-stabilizer structure, where a grey box indicates that the corresponding stabilizer (column) acts on the corresponding qubit (row). For $Z$-type stabilizers, e.g., $S_1 = Z_1Z_2Z_3Z_7$. Thinking of the columns as ancillas, note that a single error on a data qubit would leave exactly four ancillas in the $\ket{1}$ state.}
\label{fig:stabs}
\end{center}
\end{figure}

Since these all commute, by taking all possible products we can form $7$ stabilizer operators. All of these operators stabilize the logical subspace. To understand the utility of considering the additional stabilizers, it is helpful to examine which stabilizers act on which qubits, and which qubits affect which stabilizers. This is captured in Fig. \ref{fig:stabs}. From the diagram, we can point out several convenient features of the enlarged stabilizer set. In particular, note that each stabilizer acts on exactly four qubits. Also, each qubit affects the value of exactly four stabilizer values. In fact, swapping the rows and columns of Fig. \ref{fig:stabs} produces an identical structure. 

While considering only the stabilizer generators, an error on say $q_7$, will affect only a single stabilizer value ($S_1$), while an error on $q_1$ would affect all three generators.  By considering the full set we see that all qubits are on equal footing, so that without loss of generality we can examine errors on a single data qubit, knowing that the results will generalize to other qubits. 

Thus, examining $q_1$ in Fig. \ref{fig:stabs}, it is evident that any two stabilizers overlap at exactly 2 sites. For example, $S_1$ and $S_2$ both act on $q_1$ and $q_2$. This overlap property is a powerful feature of the enlarged stabilizer set, particularly in the context of measurement-based error correction.

The syndrome extraction procedure encodes the value of each stabilizer onto an ancilla qubit, requiring 14 total qubits. Here, we will label the ancillas and the stabilizers using the same symbols. If a data qubit experiences an error, then a properly extracted syndrome will leave exactly 4 ancilla qubits in the $\ket{1}$ state. Therefore, if a single extraction error occurs, corrupting a single ancilla, then we will either have 3 or 5 ancilla qubits in the $\ket{1}$ state. Without loss of generality, if $q_1$ experiences an error, and the $S_7$ value is improperly extracted, then ancilla qubits $S_1$, $S_2$, and $S_3$ will be in the $\ket{1}$ state. However, $q_1$ is the only qubit that would have caused all three of those stabilizers to be incorrectly extracted, so we deduce that $q_1$ experienced a data error.

Similarly, if 5 ancillas occupy the $\ket{1}$ state, it remains possible to uniquely identify a data qubit. Continuing with our example data error on $q_1$, suppose that in addition to the usual four stabilizers, we measure $S_4$ in the $\ket{1}$ state. Now, for qubits $q_2$ and $q_3$, we will measure 3 stabilizers indicating that they had an error. However, all 4 of the $q_1$ stabilizers indicate error. Thus, we again correctly identify the data error. We emphasize again that our example generalizes to all qubits and stabilizers. That is, \emph{no} single encoding error will cause error correction to fail. 

Lastly, if two ancilla qubits are corrupted during extraction, we will either measure 2, 4, or 6 ancillas in the $\ket{1}$ state. Clearly, we can identify 2 and 6 $\ket{1}$ stabilizers as a nonsensical syndrome. Since every pair of qubits overlaps at exactly two stabilizers, there is no way to turn a properly extracted syndrome for one qubit into a properly extracted syndrome for another qubit. Thus, we again detect a nonsensical syndrome and deduce that 2 extraction errors occurred. In this case, this syndrome is simply discarded and extracted again. 

Thus, the combination of syndrome extraction and syndrome matching is automatically robust. There is no need for ancilla verification. Furthermore, we do not need to directly entangle the ancillas with each other, and they are immune to phase errors.

For the measurement-free case, some benefits of the additional stabilizers are lost. In particular, in the measurement-based context, it is fairly straightforward to interpret syndromes with extraction errors. However, losing measurement as a resource leaves two options. On one hand, a sufficiently complicated classical logic circuit (on quantum hardware) could properly interpret a wide range of syndromes without measurement. However, we opted for an approach based on making the circuit as simple as possible, with the cost of losing some robustness of the syndrome matching.

Our procedure only corrects data errors if a syndrome is properly extracted. In the event of a faulty extraction, data qubits remain unaffected. This means that preexisting data errors can possibly survive a single error correction cycle, but our circuit prevents further harm from propagating to the data. Note that this is still sufficient for fault-tolerance, since any single-qubit error at any point in the circuit will cause logical failure. 

To do this, we use C$_4$NOT gates to correct errors on the data qubits. The syndrome is extracted onto seven ancillary qubits as in the measurement-based case. Then, the four ancillas corresponding to each data qubit serve as the controls in the C$_4$NOT gates. The full circuit is shown in Fig. \ref{fig:steane-circuit}.

\subsection*{Additional details for the Bacon-Shor code}
\label{app:bs}

The procedure for the Bacon-Shor code is quite similar to the bit-flip code. To see this, recall that the bit-flip code is comprised of logical states $\ket{000}$ and $\ket{111}$. An analogous 3-qubit phase flip code would use logical states $\ket{+++}$ and $\ket{---}$. Note that the bit-flip code, if concatenated, would require 9 data qubits. For the Bacon-Shor code, we ``concatenate'' a bit-flip code with a phase-flip code. The result is a 9 qubit code that corrects arbitrary single-qubit errors. Restricting to bit-flip errors, the Bacon-Shor stabilizer structure is identical to that of the bit-flip code. So we can apply the same method as in the bit-flip code.

For bit-flip errors, and picturing the data qubits in a $3\times 3$ array, we have stabilizers
\[
Z_U = \begin{pmatrix}
Z & Z & Z \\
Z & Z & Z \\
I & I & I
\end{pmatrix}
\]
and
\[
Z_D = \begin{pmatrix}
I & I & I \\
Z & Z & Z \\
Z & Z & Z
\end{pmatrix}.
\]

\begin{figure}[h]
\begin{center}
\includegraphics[width=.5\columnwidth]{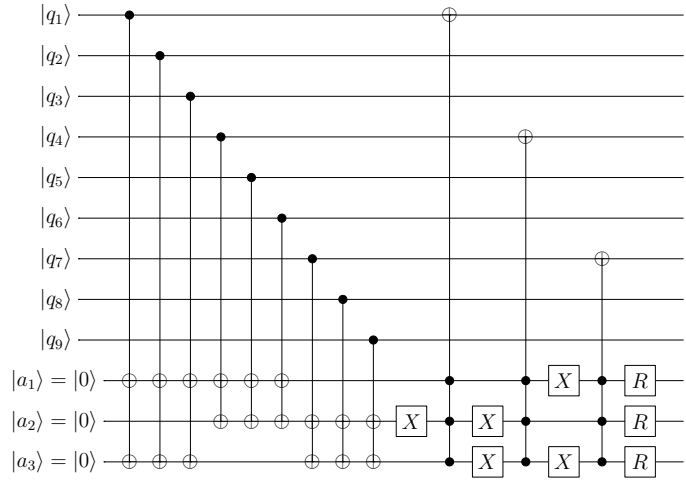}
\caption{The full bit-flip error extraction and correction circuit for the BS code. The first 9 gates are for syndrome extraction, the C$_3$NOT gates correct the errors. Phase errors are corrected in an analogous process, but data qubits are grouped by columns instead of rows. Here, the rows are $(123)$, $(456)$, and $(789)$.}
\label{fig:bs}
\end{center}
\end{figure}

Note that two $X$ operators acting in the same row commute with these stabilizers, as well as the logical operators $Z_L = Z^{\otimes 9}$ and $X_L = X^{\otimes 9}$. Thus, if a bit-flip error occurs in one row, we can correct it by applying an $X$ gate to \emph{any} qubit in the same row. With this in mind, we proceed as with the bit-flip code. We define $Z_M = Z_U\cdot Z_D$ and extract all three stabilizer values. Then, errors are corrected row by row, as opposed to qubit by qubit in the 3-qubit case. The circuit is shown in Fig. \ref{fig:bs}.

\subsection*{Simulation details}
\label{app:sim}

The error model introduces possible memory errors at every time step, and for every qubit. Thus, there are a total of $qt$ memory error sites, for $q$ total qubits and $t$ time steps. Similarly, there are a total of $g$ gate error sites. Each of these sites is affected by an error with a fixed probability, so a sequence with $i$ memory errors and $j$ gate errors occurs with probability
\begin{equation}
P(i,j) = {qt \choose i}{g \choose j} p_{\mathrm{mem}}^i (1-p_{\mathrm{mem}})^{qt-i} p_{\mathrm{gate}}^j (1-p_{\mathrm{gate}})^{g-j}.
\label{eq:prob}
\end{equation}

Here, we combine two previous simulation techniques. Note that the entire error correction procedure is Markovian. Thus, we can determine transfer rates between logical states and use these transfer rates to extract information about logical failure rates \cite{Bravyi2013}. Rather than tracking all possible states, we consider our logical states as either being logically correct, correctable, or failed. The states with correctable errors can be further split into single bit-flip, single phase-flip, and both types of error at once. Thus, we must determine the transition probabilities between these 5 subclasses of logical states. We assume that once a state has experienced a logical failure, it will never accidentally correct itself.

To determine the transfer rates between two logical states of type $a$ and $b$, we used a combinatorial expansion, similar to \cite{Wang2010}. There, the expansion was used to calculate failure probabilities directly, but we use it here to find the transition rates. That is, 
\[
T_{ab} = \sum_{i,j} \alpha_{ab} P(i,j) .
\]
We determine the $\alpha_{ab}$ by sampling over the appropriate fault paths and input states. Note that $P(i,j)$ shrinks rapidly for small error rates, so the sum can be truncated at low order. Using the computed transfer matrix, it is possible to examine the system dynamics over repeated error correction cycles. 

For a single uncorrected qubit with error rate $p$, the chance that a logical failure has occurred grows as $\sum_{i=0}^{T} p\cdot (1-p)^i$. By comparing logical failure rates with growth of this form, we can extract a logical error rate $p_L$.

\end{document}